\documentclass[aps,pra,twocolumn,amsmath,amssymb,superscriptaddress]{revtex4-2}

\usepackage[usenames,dvipsnames]{xcolor}
\usepackage{amssymb}
\usepackage{graphicx}
\usepackage{amsmath}
\usepackage{hyperref}
\usepackage{braket}
\usepackage[normalem]{ulem}
\definecolor{pink2}{rgb}{0.858, 0.188, 0.478}

\usepackage{array}
\usepackage{multirow}

\providecommand{\tabularnewline}{\\}
\usepackage{slashbox}
\begin{document}
\title{Characterization of ion-trap-induced ac-magnetic fields}
\author{Manoj K. Joshi}
\author{Milena Guevara-Bertsch}
\author{Florian Kranzl}
\author{Rainer Blatt}
\affiliation{Institut f\"ur Quantenoptik und Quanteninformation, \"Osterreichische Akademie der Wissenschaften,
Technikerstra\ss{}e 21a, 6020 Innsbruck, Austria}
\affiliation{Institut f\"ur Experimentalphysik, Universit\"at Innsbruck, Technikerstra\ss{}e 25, 6020 Innsbruck, Austria}

\author{Christian F. Roos}
\affiliation{Institut f\"ur Quantenoptik und Quanteninformation, \"Osterreichische Akademie der Wissenschaften,
Technikerstra\ss{}e 21a, 6020 Innsbruck, Austria}
\affiliation{Institut f\"ur Experimentalphysik, Universit\"at Innsbruck, Technikerstra\ss{}e 25, 6020 Innsbruck, Austria}
\email{christian.roos@uibk.ac.at}

\begin{abstract}
The oscillating magnetic field produced by unbalanced currents in radio-frequency ion traps induces transition frequency shifts and sideband transitions that can be harmful to precision spectroscopy experiments. Here, we describe a methodology, based on two-photon spectroscopy, for determining both the strength and direction of rf-induced magnetic fields without modifying any DC magnetic bias field or changing any trap RF power. The technique is readily applicable to any trapped-ion experiment featuring narrow linewidth transitions.
\end{abstract}

\maketitle
\section{Introduction} \label{sec:intro}
Improving the accuracy of atomic standards is vital for research areas investigating the variability of fundamental constants \cite{Ludlow2015, godun2014}, building frequency and time standards \cite{chou2010optical, bloom2014, Nicholson2015}, testing new theories that advocate beyond the standard model of physics \cite{Ludlow2015, Peik2021, Safronova2018}, investigating many-body dynamics \cite{Ansori2015},  and more \cite{McGrew2018, Takamoto2020}. Trapped ions are one of the ideal choices for high-precision measurements, and clock standards \cite{Brewer2019, Burt2021, Sanner2019, Gebert2015, godun2014, Guggemos2019}. Ambient electromagnetic fields limit the accuracy of trapped-ion clocks \cite{Itano2000, Arnold2018} and their systematic estimation is necessary for a reliable frequency standard \cite{Barrett2015}. One source of ambient electromagnetic fields in trapped-ion experiments is an oscillating magnetic field arising from unbalanced currents between the trap electrodes. Especially, in systems where the atomic levels are magnetically sensitive, precision measurements are affected by ac-Zeeman shifts changing the energy of electronic states due to off-resonant coupling to the trap-induced magnetic fields \cite{Beloy2023, Gan2018,hempel2014}. These shifts can cause incorrect determination of the DC-magnetic field and the Landé $g$-factor, which are estimated in experiments by taking ratios of the Zeeman spacing between various electronic states. A direct implication of such effects is in experiments where more than two electronic states are used for quantum computation, such as in universal qudit processors \cite{Ringbauer2022}. In such experiments, transition frequencies are set based on the g-factors known from the literature and an estimate of the magnetic field by measuring the splitting of Zeeman sublevels. However, offsets in estimated frequencies arise if ac-magnetic fields are present in the system. Additionally, these oscillating magnetic fields induce transitions that limit the proper compensation of the micromotion sidebands \cite{Meir2018}. Specifically, when the two effects interfere out-of-phase, the overall sideband strength may appear reduced but a trapped ion would have a considerable driven motion that is unintentionally added to balance the effects of the oscillating magnetic fields.

The strength of the oscillating magnetic fields in trapped-ion systems depends upon the unbalanced current in the trapping region and thus is proportional to the RF voltage applied for confining the ions. Previously, the effect of the rf-induced magnetic field on the atomic transition has been characterized by measuring transition level shifts as a function of the applied RF power and conversely extrapolating the ac-Zeeman shifts for the required RF powers \cite{Guggemos2019}. Complementary measurements have also been carried out by measuring the Autler-Townes splitting induced by trap-induced magnetic fields on the optical transitions \cite{Gan2018}. Both methods require changing the experimental conditions, i.e. either changing the strength of the RF field or the strength of the quantization magnetic field that is different from the normal operational value. The former method may be less precise; either due to less accurate knowledge of changes in the RF power or the occurrence of additional shifts in the transition frequency that are accompanied by ion position changes while changing the RF power. On the other hand, the latter method can impose restrictions in setting up the magnetic fields to a precise value when the field is provided by permanent magnets, which is often the preferred choice for experiments aiming for long coherence times. Here, we overcome this problem by presenting a method that does not require changing the experimental conditions and also can be directly implemented on the ions by measuring transitions that are the most sensitive to such effects. We anticipate that the new method is more precise and better suited to modern trapped ion platforms than the earlier methods.

This work focuses on characterizing both the strength and direction of a magnetic field oscillating at the RF drive frequency by examining the coupling strength on two-photon transitions. The impact of these fields on micromotion sidebands will be discussed in the context of the excess-micromotion minimization procedure, which is essential for the rf-trap-based quantum systems. We will provide insights into why it is crucial to characterize these fields for precision measurements.
The manuscript is structured as follows, in section \ref{sec:theory}, we discuss the theory of the interaction of trapped ions with the trap drive-induced electric and magnetic fields. In section \ref{sec:expmethod}), we discuss the measurement procedure and the experimental platform. In section \ref{sec:results}, we present experimental results and their analyses. 

\section{Theoretical description of the rf-induced effects}\label{sec:theory}

In RF traps, oscillating electric quadrupole potentials dynamically confine ions in an effective harmonic potential. Away from the rf-zero point/line, the rapidly varying electric fields give rise to ``excess micromotion" (EMM) of the ion \cite{Berkeland1998}. This driven motion phase modulates the laser beams whose direction is not perpendicular to the direction of the EMM. In spectra of narrow-band electronic transitions, it results in micromotion sidebands at integer multiples of the trap drive frequency. Applying RF voltages to the trap electrodes not only creates electric fields but also substantial electric currents that flow through the electrodes thus generating ac-magnetic fields at the location of ions. In a perfectly symmetric trap, the ac-magnetic fields created by currents through opposite RF electrodes would perfectly cancel each other at the rf-zero point/line. Yet, in practice, unbalanced currents often induce a magnetic field at the location of the ions. 

\begin{figure*}
    \centering
    \includegraphics[width=170mm]{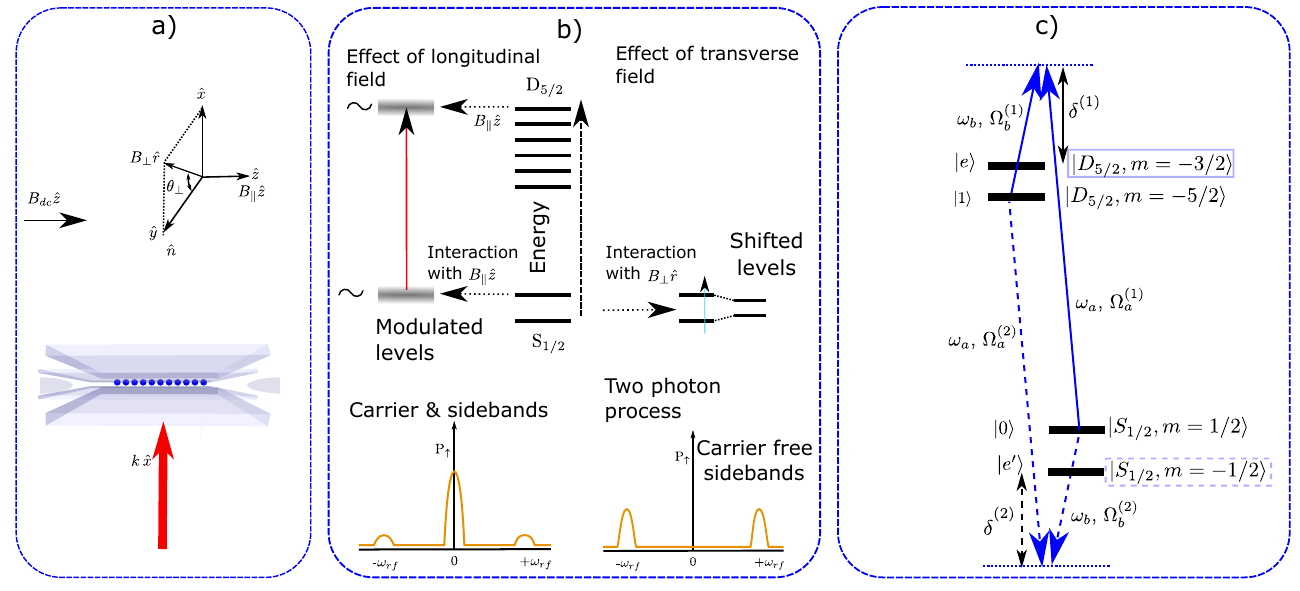}
    \caption{Illustration of AC magnetic effects in RF traps. (a) Sketch of an RF trap, magnetic field direction, and laser propagation direction. (b) The effect of the magnetically induced transitions due to longitudinal and transverse components are depicted side by side for a trapped $^{40}$Ca$^{+}$ ion. The contribution of the longitudinal component causes oscillating energy levels of the magnetically sensitive transitions, thus giving rise to sidebands around the carrier transition. The transverse component leads to two effects, (1) level shifts if the rf-magnetic coupling is off-resonant, and (2) a two-photon transition that can be induced by the ac-magnetic field in conjunction with a laser that couples two electronic levels. In the current context, the two-photon process can still produce sidebands without a carrier transition such as for $\Delta m =\pm 3$. We will call them carrier-free sidebands, which are the means of measuring the transverse component in our present studies. 
    (c) The two-photon process for an exemplary case is shown for a trapped calcium ion. Here the electronic levels $\ket{0}$ and $\ket{1}$ do not directly couple via the single photon transition. Rather, a two-photon process is involved, in which a magnetic dipole interaction couples adjacent Zeeman sublevels belonging to the same electronic state, and an electric quadrupole interaction couples levels belonging to states with different orbital angular momentum. There are two possible paths for which the ac-magnetic field mediates a coupling within the ground-state manifold and in the excited-state manifold, respectively. 
    }
    \label{fig:Schematic}
\end{figure*}

For the present discussion, we divide ac-magnetic fields into two components: First, the component parallel to the direction of the quantization magnetic field (longitudinal component), and second, perpendicular to the quantization magnetic field (transverse component), see Fig.~\ref{fig:Schematic}(a). In the theoretical treatment described below, we will initially consider the interaction of oscillating magnetic fields on Zeeman sublevels of the same total angular quantum number $J$ via magnetic dipole coupling. Here we will use the total angular momentum operator $\boldsymbol{J}$ to design the interaction Hamiltonian. Later on, when we design the Hamiltonian for a laser field that couples a pair of electronic levels with total angular quantum numbers $J$ and $J'$, the magnetic coupling between the levels will be described in terms of Pauli spin operators $\sigma^{x,y,z}$.

\subsection{Magnetic coupling in RF traps} 
The interaction Hamiltonian for an atomic ion in a magnetic field is expressed as
\begin{equation}
H = -\boldsymbol{\mu} \cdot \boldsymbol{B},    
\end{equation}
where $\boldsymbol{B} = \boldsymbol{B}_{dc} + \boldsymbol{B}_{rf}$ takes into account the contributions from the quantization field $\boldsymbol{B}_{dc}$ and the rf-current induced field $\boldsymbol{B}_{rf}$. The magnetic moment of an atom in a state with angular momentum operator $\boldsymbol{J}$ and a Land{\'e} $g$-factor  is expressed as $\boldsymbol{\mu}=-g_{j}\mu_{B}\boldsymbol{J}/\hbar$. For simplicity, $\hbar = 1$ will be assumed throughout the discussion of this paper. We decompose the oscillating magnetic field into two parts, $B_{\parallel}$  and $B_{\perp}$, that are parallel and perpendicular to the quantization field $\boldsymbol{B}_{dc} = B_{dc} \boldsymbol{\hat{z}}$. The oscillating magnetic field, expressed in terms of its longitudinal and transverse components, is given by $\boldsymbol{B}_{rf} = B_{\parallel } \cos (\omega_{rf}t) \boldsymbol{\hat{z}} + B_{\perp} \cos (\omega_{rf}t) (\sin{\theta_\perp}\boldsymbol{\hat{x}}+\cos{\theta_\perp}\boldsymbol{\hat{y}})$. Here, $\omega_{rf}$ is the frequency of the RF drive, $\boldsymbol{\hat{z}} (\boldsymbol{\hat{x}}, \boldsymbol{\hat{y}})$ denote the unit vectors and $\theta_\perp$ sets the direction of the transverse field component in the $x-y$ plane. The interaction Hamiltonian further details as
\begin{eqnarray}
\label{eq:IntHam}
H&= &g_{j}\mu_{B}B_{dc}J_{z}+g_{j}\mu_{B} B_{\parallel}J_{z}\cos(\omega_{rf}t)\\ \nonumber
&+& g_{j}\mu_{B} B_{\perp}(J_{x}\sin{\theta_\perp}+J_{y}\cos{\theta_\perp})\cos(\omega_{rf}t).
\end{eqnarray}
The above terms are responsible for three mechanisms. The first term gives rise to a Zeeman splitting. The second term modulates the energy levels at the drive frequency, producing sidebands that coincide with the micromotion sidebands (discussed further in section E) when using a laser to excite the ion on an electronic transition. The third term couples two Zeeman states via the magnetic dipole coupling. It is responsible for the ac-Zeeman shift mentioned in reference \cite{Gan2018} in the case where $\omega_{rf}$ is different than the transition frequency of adjacent Zeeman states. The term is also responsible for a two-photon transition as discussed in section D. The effect of these oscillating magnetic field components on the atomic transitions is illustrated in Fig.~\ref{fig:Schematic}b.   

\subsection{Effect of the longitudinal field on atomic transitions} \label{sec:longcompTheory}
Let us start the discussion with the first two terms in Eq.~\eqref{eq:IntHam}. In the following, we will discuss the effect of the longitudinal component of the rf-induced magnetic field on the laser-ion interaction on a narrow linewidth transition. Specifically, we will consider a pair of levels, where the ground state $|g\rangle$ has a total angular momentum $j$ and magnetic quantum number $m_j$ and the excited state $|e\rangle$ is characterized by the respective quantum numbers $(j^\prime, m_{j^\prime})$. The atomic Hamiltonian is given by $H_a = E_e\ket{e}\bra{e} + E_g\ket{g}\bra{g}$ with $E_g =-\frac{\omega_0}{2} + E_{j,m_j}$ and $E_e=\frac{\omega_0}{2} + E_{j',m_j'}$, where $\omega_0$ denotes the transition frequency between the levels in the absence of magnetic fields and 
\begin{align}\label{eq:oscElevels}
    E_{j,m_j}&=g_{j}\mu_{B}m_{j}(B_{dc}+B_{||} \sin(\omega_{rf}t))\\
    E_{j^\prime,m_{j^\prime}}&=g_{j^\prime}\mu_{B}m_{j^\prime}(B_{dc}+B_{||}\sin(\omega_{rf}t)) 
\end{align}
account for the contributions to the energies that result from the time-dependent magnetic field.

The laser-ion coupling Hamiltonian, driving transitions between $j,m_j$, and $j',mj'$ levels by coupling the ion to a travelling-wave laser field with wave vector $\mathbf{k}$, is given by 
\begin{equation}
\label{eq:HamLaserion1}
    H_{I} = \frac{\Omega}{2}(\sigma^++\sigma^-)(e^{i(\omega_lt-\mathbf{kr})}+e^{-i(\omega_lt-\mathbf{kr})}),
\end{equation}
where $\sigma^{\pm}$ is the Pauli spin raising (lowering) operator and $\Omega$ is the Rabi frequency. By moving into an interaction picture with respect to the atomic Hamiltonian $H_a$, we obtain
\begin{equation}
    \tilde{H}_{int}=\frac{\Omega}{2}[\sigma^{+}e^{-i(\delta t+ \beta\cos(\omega_{rf}t))}]
    +h.c.,
    \label{eq:HLong}
\end{equation}
where $\beta= \frac{\chi \mu_{B}  B_{||}}{\omega_{rf}}$ and $\delta=(\omega_l-\omega_0)+(g_{j}m_{j}-g_{j^\prime}m_{j^\prime})\mu_{B}B_{dc}$ is the detuning of the laser from the time-averaged atomic transition frequency. The parameter $\chi = g_{j'}m_{j'}-g_{j}m_{j}$ describes the Zeeman susceptibility, discussed previously in ref.~\cite{Meir2018}. The phase-modulation of the interaction by the time-varying longitudinal magnetic field results in sidebands to the transition at integer multiples of the rf-drive frequency, akin to the micromotional sidebands induced by an electric field at the location of the ion that oscillates at the rf-drive frequency~\cite{leibfried_wineland_2003}). The first-order sideband and the carrier coupling strengths are expressed as
\begin{eqnarray}
    \Omega_{SB}&=& \Omega J^{\pm 1}(\beta_B), \nonumber \\
    \Omega_{Carr}&=&\Omega J^{0}(\beta_B),
\end{eqnarray}
where $J^{0}(\beta_B )$ and $J^{\pm 1}(\beta_B) $ are Bessel functions of the first kind. Note that the modulation in the present case depends upon the magnetic sensitivity of the energy levels; thus we can define a modulation index 
\begin{equation}
  \beta_B=\frac{\chi \mu_{B}  B_{||}}{\omega_{rf}}.
  \label{eq:longmodindex}
\end{equation}
The modulation index $\beta_B$ can be inferred from the ratio of a sideband and the carrier transition strengths, which, in the limit of weak modulation, is given by $\beta_B \approx 2\Omega_{SB}/\Omega_{Carr}$. For the estimation of the modulation index, one can thus drive Rabi oscillations on the carrier and blue sideband of a narrow linewidth transition and evaluate the Rabi frequencies and their ratio.

\subsection{Effect of the transverse component on atomic transitions}
For the transverse component, we first define the raising and lowering operators $J_{\pm}=(J_{x}\pm i J_{y})$. We use them to express the last term of Eq.~\ref{eq:IntHam} as
\begin{equation}
H_{I}= \frac{1}{2} \mu_{B}B_{\perp}g_{j} (J_{+}e^{-i\theta_\perp}+J_{-} e^{i\theta_\perp})\cos(\omega_{rf}t).
\end{equation}
Here, the operator $J_{\pm}$, gives rise to a coupling of $m\leftrightarrow m\pm1$ transitions,  which represent spin flips between the Zeeman sublevels within a single fine structure level. For the current discussion, we will restrict the derivation to a pair of adjacent levels, $m$ and $m^\prime = m\pm 1$, and rewrite the Hamiltonian in $\sigma_\pm$ instead of $J_\pm$. In this new notation, the Hamiltonian is written as 
\begin{equation}
H_{I}= \frac{1}{2}\Omega_{m,m'}(\sigma_{+}e^{-i\theta_\perp}+\sigma_{-} e^{i\theta_\perp})\cos(\omega_{rf}t),
\end{equation}
where the Rabi frequency is expressed as
\begin{equation}
\label{eq:magneticRabi}
   \Omega_{m,m'}=\frac{1}{2}\mu_{B}B_{\perp}g_{j}\sqrt{j(j+1)-m(m\pm1)}.  
\end{equation}
The $\pm$ sign indicates that the spin-flip is accompanied by $\Delta m=\pm 1$. Now let us reformulate this Hamiltonian in the interaction picture of the atom, assuming that the Zeeman levels are spaced by frequency $\omega_{m,m'}$, such that 
\begin{eqnarray}
\label{eq:12}
    \tilde{H}_{int}&=&\frac{1}{2}\Omega_{m,m'}[\sigma_{+} e^{-i\theta_\perp}(e^{i(\omega_{m,m'}+\omega_{rf})t }+e^{i(\omega_{m,m'}-\omega_{rf})t})] \nonumber \\ 
    &+& h.c..
\end{eqnarray}
After making a rotating wave approximation the resultant Hamiltonian simplifies to
\begin{equation}
\tilde{H}_{int}=\frac{1}{2}\Omega_{m,m'}(\sigma_{+}e^{-i\delta_m t-i\theta_\perp}+\sigma_{-}e^{i\delta_m t+i\theta_\perp}),    
\end{equation}
where $\delta_m (=\omega_{rf} -\omega_{m,m'} $). For a resonant case (i.e. $\delta_m=0$), the population transfer takes place at a rate given by the Rabi frequency $\Omega_{m,m'}$. For the far-detuned case, the drive field shifts the energy levels through the ac-Zeeman effect \cite{Gan2018}, which is expressed as
\begin{equation}
    \delta_{ac} =  -(g_j \mu_B B_{\perp})^2 \frac{m}{8} (\frac{1}{\omega_{rf}-\omega_{m,m'}}-\frac{1}{\omega_{rf}+\omega_{m,m'}}).
\end{equation}
Note that the expression contains the co- and counter-rotating terms of the field as for the present studies they are comparable in amplitude. This ac-Zeeman shift, if not considered correctly, can lead to errors in the estimation of the Landé g-factor. The $B_{\perp}$ term has been measured previously in RF traps via the measurement of the Autler-Townes splitting as discussed in ref. \cite{Gan2018}, requiring setting the quantization field such that a resonant coupling via magnetic fields is achieved. 
In our experiments, instead of using a resonant magnetic dipole coupling, we will combine an off-resonant coupling with the optical light field that is tuned to achieve a resonant two-photon transition, therefore, we don't have to change any experimental conditions. This approach will be presented in more detail in the next section.

\subsection{Two-photon transitions} 
The transverse component of the ac-magnetic field, in combination with a laser coupling two electronic levels, can give rise to two-photon transitions. In the scenario sketched in Fig.~\ref{fig:Schematic}c for an electric quadrupole transition in Ca$^+$, RF photons off-resonantly couple the levels $\ket{1}$ and $\ket{e}$ via a magnetic dipole coupling as discussed in the previous section. Simultaneously, the laser photons couple the levels $\ket{0}$ and level  $\ket{e}$ via electric quadrupole transition rules. Note, that both fields are assumed to be detuned by an amount $\delta$ from the mediator level $\ket{e}$ such that the two-photon transition manifests a population transfer from level $\ket{0}$ to $\ket{1}$ while only virtually populating the level $\ket{e}$. For achieving a resonant two-photon process, the laser frequency $\omega_a$ needs to be tuned such that $\omega_a=\omega_{01}+\omega_{b}$. For the current work, $\omega_{b}$ is set equal to the RF frequency ($\omega_{rf}$), and only laser frequency $\omega_a$ is tuned such that resonant two-photon transitions are observed. Throughout the manuscript $\omega_{ij}$ will represent the frequency spacing of levels indexed by $i$ and $j$.  

The two-photon Rabi frequency is defined as the product of the Rabi frequencies of the RF and optical photons to their respective transitions and divided by the detuning \cite{Bateman2010}. In the situation sketched in Fig.~\ref{fig:Schematic} (c), there is a second path linking the levels $\ket{0}$ and $\ket{1}$ via virtually exciting the energy level $\ket{e'}$, thus the effective Rabi frequency for the two-photon transition is expressed as (see Appendix for a detailed derivation)
\begin{equation}\label{eq:twophoton}
\Omega_T =  \frac{\Omega_a^{(1)} \Omega_{b}^{(1)}}{2 \delta^{(1)}} -\frac{\Omega_a^{(2)} \Omega_{b}^{(2)}}{2 \delta^{(2)}},
\end{equation}
$\delta^{(i)}$ is the detuning of the two photons from their respective transitions, where $i$ is 1 or 2 representing two paths. $\Omega_a^{(i)}$ and $\Omega_{b}^{(i)}$ are Rabi frequencies of the optical and RF transitions, respectively. The negative sign of the second term accounts for a destructive interference of the two paths linking the levels.

Note that a resonant two-photon transition between $\ket{0}$ and $\ket{1}$ could also be achieved by setting the laser frequency to $\omega_a=\omega_{01}-\omega_{b}$. In this case, the second leg of the transition would be provided by the counter-rotating terms of the coupling to the magnetic dipole transition and one would have to replace the detunings $\delta^{(i)}$ in eq.~(\ref{eq:twophoton}) by 
$\delta^{(1)} = \omega_a-\omega_{0e}$ and $\delta^{(2)} = \omega_a-\omega_{1e'}$.

\subsection{Micromotion sidebands}\label{sec:EMM}
The phase-modulation of magnetic-field sensitive atomic transitions induced by longitudinal ac-magnetic fields results in sidebands that need to be distinguished from the micromotional sidebands of an ion displaced from the rf-zero point/line \cite{Meir2018}, which occur at the same laser detuning. 
Similarly, as driving two-photon resonances enabled by the trap's ac-magnetic field requires the laser frequency to be set to a value for which also first-order micromotional sidebands can be observed, excess micromotion \cite{Berkeland1998} has to be included in a treatment of the effects of ac-magnetic fields.

If we disregard the secular motion and consider that the ion is shifted away from the RF null point due to some stray electric fields \cite{Berkeland1998}, the ion motion is purely described by driven motion due to the rf-field. Following the treatment given in Ref.~\cite{Meir2018}, the micromotion $\bold{u_{\text{EMM}}}$ can be described in terms of the components that are either in phase ($u_{I}$) or phase-shifted by $\pi/2$ ($u_Q$) with respect to the RF drive field; i.e. $\bold{u_{\text{EMM}}} = \sum_{i=x,y,z} \big[u^i_{I} \cos{\omega_{rf} t} +u^i_{Q} \sin{\omega_{rf} t}\big] \hat{i}$. These two components imprint a phase modulation on the laser beam in the ion's rest frame. The relevant modulation index is written as $\bold{k}\cdot\bold{u_{EMM}}$, where $\bold{k}$ is the wavevector of the laser beam and $\hat{i}$ represents $\hat{x},\hat{y},\hat{z}$. 
The in-phase contribution of EMM is caused by stray fields, pushing an ion away from the rf-null line, thus the corresponding modulation index will depend upon the strength of the stray potentials at the ions. On the other hand, a phase shifted term $u_Q$ can result from a phase imbalance of the drive fields applied to the rf-electrodes~\cite{Berkeland1998}. 

By moving into an interaction picture with respect to the atomic Hamiltonian $H_a$, we can write the interaction Hamiltonian for micromotion (retaining an expression similar to Eq.~\ref{eq:HLong})
\begin{equation}
    \tilde{H}^{(M)}_{int} = \frac{\Omega}{2} [\sigma^+e^{-i(\delta t +\beta_{I}\cos(\omega_{rf}t)+\beta_{Q}\sin(\omega_{rf}t)}+h.c.],
    \label{eq:HEmm}
\end{equation}
where $\Omega$ is the carrier Rabi frequency, and the modulation indices are expressed as
\begin{eqnarray}
\label{eq:modulationindexE}
\beta_{I} = \bold{k}\cdot (u_{I}^x \hat{x} +u_{I}^y \hat{y}+u_{I}^z \hat{z}),\nonumber \\ 
\beta_{Q} =  \bold{k}\cdot (u_{Q}^x \hat{x} +u_{Q}^y \hat{y}+u_{Q}^z \hat{z}).
\end{eqnarray}
Notably, after making a rotating wave approximation to Eq.~\ref{eq:HEmm}, we reach the similar derivation discussed in Section B, and henceforth we can derive the sideband strengths as a function of the corresponding modulation indices. The EMM sidebands appear at the same frequency as the transitions caused by the longitudinal and transverse components of the AC magnetic field. More specifically, we model our trapping device as capacitively loaded. This means that the electric field experienced by ions is phase-shifted by $\pi/2$ with respect to the electric current flowing through the trap electrodes. From the earlier discussion of the longitudinal component of the magnetic field (see Sec.\ref{sec:longcompTheory}) and EMM (Sec.~\ref{sec:EMM}), we can argue that the modulation caused longitudinal component of the AC magnetic field and in-phase EMM will coherently add up and can be treated together, as already done in Ref.~\cite{Meir2018}.

\subsection{Extraction of magnetic field components} In this section, we explain how to estimate the longitudinal and transverse components of the magnetic fields. As discussed in the previous sections, all three effects; i.e. modulation by EMM, modulation by longitudinal component of the AC magnetic field, and two-photon effect induce transitions at the same frequency. However, these effects can be easily segregated with a specific choice of transitions and probing schemes. For the estimation of the longitudinal component, we can design a scheme where any two-photon transitions are suppressed but single-photon transitions are allowed. In a multi-level atomic system, a specific choice of laser polarisation, and direction of the quantization field can facilitate this situation. On the other hand, for two-photon transitions, we can choose the scheme where the two-photon transitions are present and the coupling strength for a single-photon vanishes. For example, as shown in Fig.~\ref{fig:Schematic})c, one of the appropriate transitions is $\Delta m =\pm 3$, which prohibits single-photon transitions but a two-photon transition is still feasible. See Sec.~\ref{sec:expmethod} for a detailed discussion on the choice of specific transitions. Below, we detail the expressions for both components of the magnetic fields simplified for the current measurement scheme. 

Now, let us first discuss how to measure the longitudinal component. A combined modulation index due to EMM and $B_{||}$  can be evaluated from Eq.~\ref{eq:modulationindexE} and \ref{eq:longmodindex}, which is expressed as 
\begin{equation}
\label{eq:betachiedc}
    \beta = \sqrt{\biggl(\beta_{I}+ \frac{\mu_{B}B_{||} \chi}{\omega_{rf}}\biggr)^2+\beta_{Q}^2}.
\end{equation}
The combined modulation index can be measured by evaluating the ratio of the sideband and the carrier Rabi frequencies, i.e. $\beta=2\Omega_{SB}/\Omega_{Carr}$ and can be later used to estimate the strength of the longitudinal component. Here we can reason that sideband strength minimization could be achieved when the left term in Eq. \ref{eq:betachiedc} is zero. The equation also implies that the compensation voltages for minimum micromotion will vary with the transitions that are being probed. Eq.~\ref{eq:betachiedc} can be seen as presenting a hyperbolic dependence of $\beta$ to the applied compensation voltage $U$, hence we denote it as a phenomenological expression $\beta_U =\sqrt{A^2(U_0+U_\chi-U)^2+\beta_Q^2}$ used previously in Ref.\cite{hempel2014}. Here,  the compensation voltage $U$ is adjusted in experiments to minimize the micromotion modulation index. $U_0$ is a stray potential and $U_\chi$ is a $\chi$-dependent term in Eq.~\ref{eq:betachiedc}. $A$ is a coefficient that controls the slope of $\beta_U(U)$ in the regime of large micromotion (i.e. $|U-U_0-U_\chi|\gg 0 $).  From here the longitudinal component can be expressed as    
\begin{equation}
 B_{||} = \frac{\omega_{rf}}{ \mu_{B}} \frac{d(AU_\chi)}{d\chi}. 
  \label{eq:longBComp}
\end{equation}

Now let us consider the second situation where a single-photon transition is forbidden but a two-photon transition is allowed. For our current discussion, the transverse component drives two-photon transitions and hence can be evaluated independently of the longitudinal component. Thus, the transverse component can be estimated by measuring the two-photon coupling strength in the experiment. An expression for $B_{\perp}$ can be derived from Eq.~\ref{eq:magneticRabi} and Eq.~\ref{eq:twophoton};
\begin{equation}
 B_{\perp} = \frac{2\Omega_T}{\sum_i (\mu_B g_{j_i} \sqrt{j_i(j_i+1)-m_i(m_i\pm1)}) c_i \Omega^{(i)}/\delta^{(i)}}, 
  \label{eq:transBcomp}
\end{equation}
where $i$  implies the paths involved in the two-photon transitions and $c_i=1$ or -1 depending upon the path taken during the two-photon transition. 

The expressions for the longitudinal and transverse components are specifically developed to suit the methodology of measuring them in our experimental setup. Previous studies have dealt with only one of the two components of the AC magnetic field at a time, i.e., either the longitudinal component in Ref.~\cite{Meir2018} or the transverse component in Ref.~\cite{Gan2018}. Our manuscript presents the first joint investigation of both components and their measurements in the same trapping device.   

\section{Experimental method}\label{sec:expmethod}
Experiments presented in this manuscript are performed on two separate macroscopic linear Paul traps, the first dedicated to quantum simulation experiments (QSIM trap) \cite{hempel2014} and the second one to precision measurements (precision trap) \cite{Guggemos2019}. Both traps have the same physical dimensions but are operated at different rf powers, hence different radial confinement frequencies. The QSIM trap, essentially used for storing long ion strings, is operated at 2.93~MHz radial center of mass mode (COM) frequency. The precision trap is mostly used to store single ions and its highest COM mode secular frequency is 1.873~MHz. The QSIM and the precision traps are operated at 29.687 MHz and 32.351 MHz rf-drive frequency, respectively. Detailed descriptions of the two experimental setups can be found in ref. \cite{hempel2014, Guggemos2019}. For the present discussion, the trapping conditions are not extremely important but they will allow us to qualitatively understand the measured magnetic field components. 

$^{40}$Ca$^+$ ions are laser-cooled using Doppler cooling, polarization-gradient cooling, and resolved-sideband cooling for the case of a linear ion chain ($N=8$) in the QSIM trap \cite{Kranzl2022, Joshi2020}. In the precision trap, a single ion is cooled via Doppler and resolved-sideband cooling. A narrow-linewidth 729 nm laser beam is used to coherently drive the manifolds of $\mathrm{S}_{1/2} \leftrightarrow \mathrm{D}_{5/2}$ transitions, from a direction perpendicular to the RF null line. Ions are detected by both photo-multiplier tubes and electron-multiplying CCD cameras to enable spatially resolved fluorescence detection. A 4.17~(3.05) Gauss magnetic field oriented along the traps' RF zero line is used for the generation of the quantization fields in the QSIM (precision) trap.  

For the measurements of the RF magnetic field, we probe specific transitions that allow us to distinguish the effects of EMM and the longitudinal/transverse magnetic field components. For example, the transverse component can be probed by exciting transitions that do not directly couple via the quadrupole transition rules but can be driven by a two-photon process, see Fig.~\ref{fig:Schematic}(b) and \ref{fig:Schematic}(c). Similarly, the longitudinal component can be probed on magnetic-field sensitive transitions that have sidebands due to oscillating magnetic fields but vanishing two-photon transition coupling strengths. In this context, it is important that we probe several transitions that show different sensitivities to the magnetic fields so that we can separate the effect of EMM and longitudinal components. This lets us probe the two magnetic-field components independently and also distinguish them from the effects of EMM. Note that the main figures and data presented here are mainly shown for the QSIM trap. However, for comparison, we also discuss the final results of the measurement of transverse and longitudinal components in the precision trap. 

\section{Results and discussion}\label{sec:results}
\subsection{Measurements of the transverse component}
\begin{figure}
\includegraphics[scale=1]{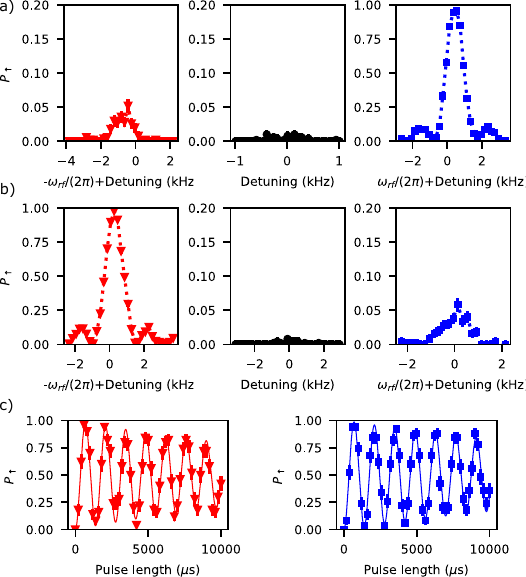}
\caption{Two-photon resonances detected in a trapped calcium ion when driving the ground and excited state manifolds with a 729 nm laser, leading to a change of the magnetic quantum number by $\Delta m =-3$ (a) and  $\Delta m =+3$ (b). The panels show the main transitions and their sidebands in the QSIM trap. In (a,b), dashed lines are a guide to the eye. (c) Excitation of Rabi oscillations on the two strongest transitions appearing in (a) and (b). The solid lines are fit to sinusoidal functions with damping. Experimental data points contain the error bars which are estimated from the quantum projection noise model and their sizes are comparable in size to the symbols. }
\label{fig:TransFig2}
\end{figure}

Transverse components are measured by probing the $|$S$_{1/2},m=+1/2\rangle \leftrightarrow |$D$_{5/2},m=-5/2\rangle$ and $|$S$_{1/2},m=-1/2\rangle \leftrightarrow |$D$_{5/2},m=+5/2\rangle$ transitions. As the magnetic quantum number changes by $\Delta m=\pm 3$, the transition cannot be excited by absorption of a single photon but is accessible by a two-photon process (involving an optical and an RF photon) when the laser is detuned by $\pm \omega_{rf}$ from the two-photon transition frequency.  The corresponding spectra are shown in Fig.~\ref{fig:TransFig2} (a) and (b). Here, the black data points indicate no excitation for the case when the laser frequency matches the two-photon transition frequency. Data shown with red triangles and blue squares display excitation peaks when the laser is detuned by an amount equal to the RF drive frequency $\omega_{rf}$. The sidebands differ in height as the detuning from the electronic state mediating the transition depends on whether the laser couples to the lower or the upper sideband.

\begin{table*}
\caption{Coupling strengths for a transverse component of the AC magnetic field and parameters for transitions. }\label{tb:tableCoupleStrngth}
\begin{tabular}{|c|c|c|c|c|c|c|c|c|c|}
\hline 
Transition & Sidebands & $\delta^{(1)}/2\pi$(MHz) & $\delta^{(2)}/2\pi$(MHz) & $C_{f}^{(1)}$ & $C_{f}^{(2)}$ & M$^{(1)}$- $D_{5/2}$  & M$^{(2)}$- $S_{1/2}$ & $\Omega_{T}/2\pi$(Hz) & $B_{rf}$(mG)\tabularnewline
\hline 
\hline 
\multirow{2}{*}{$\Delta m=-3$}& RSB & 36.69 & 41.372 & $\frac{1}{\sqrt{5}}\times0.515$ & $1\times0.515$ & $m=-\frac{3}{2}$ & $m=-\frac{1}{2}$ & 105(3)  & 283(8) \tabularnewline
\cline{2-10} 
 & BSB & 22.68 & 17.996 & $\frac{1}{\sqrt{5}}\times0.986$ & $1\times0.986$ & $m=-\frac{3}{2}$ & $m=-\frac{1}{2}$ & 715(6) & 270(2) \tabularnewline
\hline 
\multirow{2}{*}{$\Delta m=+3$} & RSB & 22.68 & 17.996 & $\frac{1}{\sqrt{5}}\times0.979$ & $1\times0.979$ & $m=+\frac{3}{2}$ & $m=+\frac{1}{2}$ & 723(8) &   275(3)   \tabularnewline
\cline{2-10} 
 & RSB & 36.69 & 41.372 & $\frac{1}{\sqrt{5}}\times0.566$ & $1\times0.566$ & $m=+\frac{3}{2}$ & $m=+\frac{1}{2}$ & 112(2)  &  274(5) \tabularnewline
\hline 
\end{tabular}
\end{table*}

The coherent excitation of the two strongest sideband transitions is shown in Fig.~\ref{fig:TransFig2}(c). The measured two-photon Rabi frequencies $\Omega_T$ are listed in Table~\ref{tb:tableCoupleStrngth}. In order to extract the strength of the transverse ac-magnetic field component from the data, knowledge of the laser-ion coupling strengths is mandatory. For this, a Rabi frequency of $\Omega_{0}= 2\pi\times 131.7 (\pm 0.7)$ kHz was measured when the laser resonantly excited the $\ket{S_{1/2},m=+1/2}$ $\leftrightarrow$ $\ket{D_{5/2}, m=+5/2}$ transition. In combination with a measurement of the diffraction efficiency of the acousto-optical modulator (AOM) used for frequency-shifting the laser, these measurements enable an estimation of the Rabi frequencies $\Omega_a^{(i)}$ (cf. eq.~(\ref{eq:twophoton}) and Fig.~\ref{fig:Schematic} c) entering the calculation of $\Omega_T$. We express the Rabi frequency as $\Omega_a^{(i)} = C_f^{(i)} \Omega_0$, where $C_f^{(i)}$ is a prefactor that includes the Clebsch-Gordan coefficients and the variation of the diffraction efficiency of the AOM over the applied frequency range of the AOM. Coefficients $C_f^{(i)}$ and detunings used in Eq.~\eqref{eq:twophoton} are presented in Table~\ref{tb:tableCoupleStrngth}. 

Using Eq.~(\ref{eq:transBcomp}), the average value of the transverse component $B_\perp$ is measured to be 276 (3) mG in the QSIM trap. In the precision trap, the value comes to be about 67(9) mG. The values differ between the two traps by a factor of 3.6. The major part of this discrepancy can be explained by their difference in the secular frequencies and the expected trap capacitance, however, the exact reason for the difference in the AC magnetic field is still unclear. 

\subsection{Measurements of the longitudinal component}\label{subsec:longcomp}
\begin{figure}
\includegraphics[scale=1]{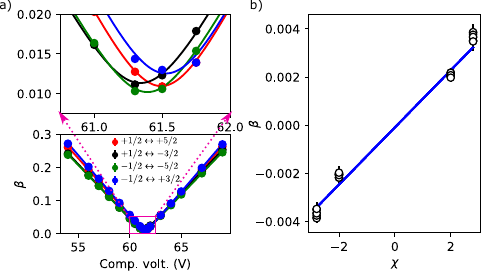}
\caption{(a) The measured modulation index of trap-drive-induced sidebands as a function of compensation voltage for four transitions in the QSIM trap (inset: zoomed around the minimum). Solid lines are fitted to a hyperbolic function, $\beta_U = A\sqrt{(U_0+ U_\chi -U)^2+\beta_Q^2}$. (b) Modulation index as a function of atomic susceptibility (squares are experimental data points and the solid line is a linear fit). Experimental data points contain error bars which are not invisible in some cases due to large marker sizes. } 
\label{fig:LongComp}
\end{figure}

We probe the four transitions between the S$_{ 1/2}$ and D$_{5/2}$ manifolds with $\Delta m = \pm 2$. We set the laser beam polarization and propagation to be perpendicular to the dc-magnetic field in order to suppress any two-photon excitation between the involved levels. For example, this setting of the laser field does not couple to any single-photon $\Delta_m =\pm1$ transitions that would have led to two-photon transitions on the $\Delta m = \pm 2$ transitions at the same frequency of the micromotion sidebands. We compensate for micromotion by changing the voltage applied to a compensation electrode of the ion trap while probing the blue micromotion sideband of the above-mentioned transitions; see Fig.~\ref{fig:LongComp}(a). Here, depending on the magnetic susceptibility of the transitions, the $B_{||}$ component shifts the voltage at which the hyperbola takes on its minimum value. The voltages minimizing micromotion for the two magnetically most sensitive transitions differ by about 200 mV. This differential shift in compensation voltage corresponds to a difference in the modulation indices $\beta \approx 0.007$. In Fig.~\ref{fig:LongComp}(b), we show a linear fit to the estimated modulation indices for all four transitions. After solving Eq.~\eqref{eq:longBComp}, we find the longitudinal component strength to be 25.2(5) mG in the QSIM trap. A similar set of measurements is carried out on the precision trap where the longitudinal component is measured to be 9(8)~mG.

\subsection{Combined AC magnetic fields and micromotion contributions}
The measurement of the transverse field component presented up to now does not provide any information about the direction of the component within the plane that is normal to the quantization axis. As the direction of the transverse component, $\boldsymbol{B}_\perp$ determines the phase factor of the two-photon coupling matrix element, additional information about the RF field direction can be obtained by measuring the Rabi frequency of those electronic transitions where the two-photon excitation process competes with a single-photon excitation process. 

\begin{figure}
    \centering
    \vspace{0.5cm}
\includegraphics[scale=1]{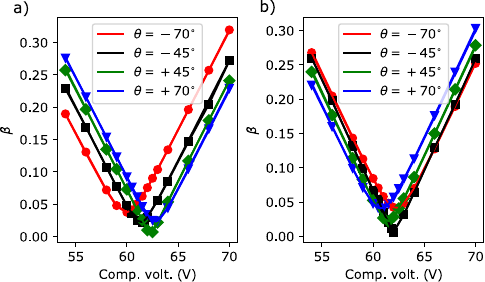}
    \caption{Modulation index ($\beta$) as a function of compensation voltage in the QSIM trap; (a) for red micromotion sideband of the $\ket{S_{1/2},m= + 1/2} \leftrightarrow \ket{D_{5/2},m=+ 5/2}$ transition and (b) blue micromotion sideband of the  $\ket{S_{1/2},m= - 1/2} \leftrightarrow \ket{D_{5/2},m=- 5/2}$  transitions. Here we perform measurements with four polarization angles -70 (red discs), -45 (black squares), +45 (green diamonds), and +70 (blue triangles) of the light field. The solid lines are the least-squares fits to the experimental data. Error bars are smaller than the markers and hence are not visible.}
    \label{fig:combinedEffect}
\end{figure}

For a more comprehensive study, we therefore probe transitions that are simultaneously sensitive to EMM, $B_\perp$, and $B_{||}$. Particularly, the sidebands which are detuned by $\mp \omega_{rf}$ from the $\ket{S_{1/2},m=\pm 1/2} \leftrightarrow \ket{D_{5/2},m=\pm 5/2}$ transition are appropriate to probe all three contributions. We reiterate the micromotion minimization procedure for these sidebands for four values of laser polarisation angles. In Fig.~\ref{fig:combinedEffect}, the data points in red discs, black squares, green diamonds, and blue triangles correspond to the polarization angles ($\theta$), -70, -45, +45, and +70 degrees,  respectively. The polarization angles are referenced to a unit vector~$\boldsymbol{\hat{n}}$ that is perpendicular to the laser propagation direction~$\boldsymbol{\hat{k}}$ and the quantization axis~$\boldsymbol{\hat{z}}$.

In Fig.~\ref{fig:combinedEffect}(a), we plot the modulation index, measured on the red rf-sideband of the $\ket{S_{1/2},m= + 1/2} \leftrightarrow \ket{D_{5/2},m=+ 5/2}$ transition, as a function of the applied compensation voltage. Here, the $B_\perp$ component contributes to the two-photon transition, which is mediated by the $\ket{D_{5/2},m=+ 3/2}$ level and its coupling depends on the polarization angle of the light field. The other two contributions, the EMM and $B_{||}$, together imprint phase modulation onto the light field and manifest sidebands on the spectra. 

Two observations are made from the experimental data; (1) The compensation voltage $V_{min}$ for achieving minimum modulation index depends on the angle $\theta$. (2) The minimum modulation index $\beta_{min}$ varies as a function of $\theta$ and the hyperbolic curves shift asymmetrically when $\theta$ flips the sign. For the first observation, the two-photon process competes with EMM depending on the angle $\theta$, and this changes $V_{min}$ corresponding to the minimum value of the modulation index. The second observation is only explainable if we take into account an angle between $B_\perp$ and a unit vector~$\boldsymbol{\hat{n}}$. Notably, this angle accounts for a phase offset between the two-photon process and the modulation due to EMM. A similar observation is made for the blue sideband of the $\ket{S_{1/2},m= - 1/2} \leftrightarrow \ket{D_{5/2},m=- 5/2}$ transition, see Fig.~\ref{fig:combinedEffect}(b). In order to understand these effects further, we carried out numerical simulations using a time-dependent Schr\"odinger equation (TDSE) on QuTiP, and the results are presented in the Appendix.

A quantitative assessment of the experimental data is carried out by performing a least-squares fitting of the data with the numerically solved TDSE. Here we extract the angle $\theta_{B_\perp}$ between the $B_{\perp}$ and $\boldsymbol{\hat{n}}$, and the RF phase mismatch $\phi_{rf}$. In Fig.~\ref{fig:combinedEffect}, solid curves are the fits. The angle $\theta_{B_\perp}$ comes to be $-2.5(3)$ radians by fitting $\Delta m = +2$ transition and $-2.2(3)$ radians by fitting $\Delta m = -2$ transition, which match within the error bars. The phase $\phi_{rf}$ in terms of modulation index is estimated to be $\beta_{Q} = 0.014(1)$ and $\beta_{Q} = 0.012(9)$ for the above two cases, respectively. The $\beta_{Q}$ value is also consistent with previous measurements of the micromotion modulation index carried out in the same trapping system \cite{hempel2014}. The equivalent phase $\phi_{rf}$ comes to be $ 5(3) \times 10^{-5} $ radians. In our fitting procedure, the parameters corresponding to the strength of $B$ field components and the laser-ion coupling to the various transitions are inserted based on the results obtained in previous subsections where the individual contributions were measured directly with the trapped ions.

\section{Conclusion} \label{sec:conclusion}
In this article, we characterized the effects of oscillating magnetic fields, produced by an unbalanced current flowing between the trap electrodes, on trapped ions. We measured the strength of transitions driven by the transverse component of these fields via a two-photon excitation method. By probing transitions for which $\Delta m =\pm 3$, we were able to measure the transverse ac-magnetic field strength without changing the experimental conditions as in previous works \cite{Guggemos2019, Gan2018}. The strength of the longitudinal ac-magnetic field component was measured by quantifying the differential shift in micromotion compensation when probing various transitions with different magnetic sensitivity, as previously done in ref. \cite{Meir2018}. 

We demonstrated that additional effects arise if the coupling between a pair of levels is enabled by both longitudinal and transverse ac-magnetic fields. Interference of such coupling processes enables a determination of the ac-magnetic field's direction. In addition, it can give rise to ''micromotion'' sidebands that cannot be completely suppressed by changing the voltage on electrodes used for compensating micromotion. Therefore, the minimal achievable micromotion modulation index cannot be directly attributed to undesired phase shifts of the RF-voltage applied to different rf-electrodes as it could also arise from the presence of ac-magnetic fields.  

Our work contributes to a better understanding of ac-magnetic fields that can give rise to a couple of undesired effects: improper micromotion compensation by minimization of the modulation index on an RF-sideband of a single transition can lead to excess micromotion resulting in transition frequency shifts caused by second-order Doppler and ac-Stark shifts that are harmful in experiments where trapped ions are employed for constructing optical frequency standards. Ac-magnetic frequency shifts resulting from the off-resonant coupling of magnetic dipole transitions by trap-induced ac-magnetic fields give rise to further transition frequency shifts that can matter in quantum computation and simulation experiments with qudits where the uncharacterized transition frequency shifts on multiple electronic levels give rise to gate errors \cite{Ringbauer2022, Hrmo2023}.

\begin{acknowledgements}
The research work presented here has received funding under Horizon Europe programme HORIZON-CL4-2022-QUANTUM-02-SGA via the project 101113690 (PASQuanS2.1) and Institut f\"ur Quanteninformation GmbH.  
\end{acknowledgements}

\appendix
\section{Two-photon transitions}
Below we derive an expression for the lambda ($\wedge$-type) and vee ($\vee$-type) transitions simultaneously taking place in a 4-level atom. A schematic of the transitions is presented in Fig.~\ref{fig:Schematic}c. The primary levels are denoted by $\ket{0}$ and $\ket{1}$. The virtual levels are denoted $\ket{e}$ and $\ket{{e^{\prime}}}$. The respective eigenenergies of these states are $\omega_0$, $\omega_1$, $\omega_e$ and $\omega_{e^\prime}$. The atomic Hamiltonian is expressed as
\begin{equation}
    H_a=\omega_0\ket{0}\bra{0}+\omega_1\ket{1}\bra{1}+\omega_e\ket{e}\bra{e}+\omega_{e^{\prime}}\ket{e^{\prime}}\bra{e^{\prime}}.
\end{equation}
As sketched in the figure, we will consider that both photons are detuned from the respective transition levels by the same amount. We will denote the two paths by superscript $(1)$ and $(2)$. Photon frequencies will be denoted by $\omega_a$ and  $\omega_b$. Corresponding Rabi frequencies are denoted by $\Omega_a^{(i)}$ and $\Omega_b^{(i)}$. We will be assuming a rotating frame in which state $\ket{e}$ and $\ket{{e^{\prime}}}$ are rotating with frequency $\delta^{(1)}$ and $\delta^{(2)}$. Let us write down the Hamiltonian describing the interaction of two photons with a 4-level atom
\begin{widetext}
\begin{eqnarray}
\begin{aligned}
H_{I}= & \frac{\Omega_a^{(1)}}{2}\big[\ket{0}\bra{e}+\ket{e}\bra{0}\big](e^{i \omega_a t}+e^{-i \omega_a t}) 
 +\frac{\Omega_b^{(1)}}{2}\big[\ket{1}\bra{e}+\ket{e}\bra{1}\big](e^{i \omega_b t}+e^{-i \omega_b t}) \\ 
& +\frac{\Omega_a^{(2)}}{2}\big[\ket{1}\bra{e^{\prime}}+\ket{e^{\prime}}\bra{1}\big](e^{i \omega_a t}+e^{-i \omega_a t})  
 +\frac{\Omega_b^{(2)}}{2}\big[\ket{0}\bra{e^{\prime}}+\ket{e^{\prime}}\bra{0}\big](e^{i \omega_b t}+e^{-i \omega_b t}).  
\end{aligned}
\end{eqnarray}
\end{widetext}

Let us define a free Hamiltonian $H_0= H_a+ H_{rot}$ such that the interaction Hamiltonian in the interaction picture with respect to $H_0$ turns into a time-independent Hamiltonian. Here 
\begin{eqnarray}
    H_{rot} =     \delta^{(1)} \ket{e}\bra{e} - \delta^{(2)} \ket{{e^{\prime}}}\bra{{e^{\prime}}}.
\end{eqnarray}
Thus $H_0$ becomes 
\begin{eqnarray}
\begin{aligned}
H_{0} &=\omega_0 \ket{0}\bra{0} +\omega_1 \ket{1}\bra{1} \\
&+ (\omega_e +\delta^{(1)}) \ket{e}\bra{e} + (\omega_{e^{\prime}} - \delta^{(2)}) \ket{{e^{\prime}}}\bra{{e^{\prime}}}.
\end{aligned}
\end{eqnarray}
The total system Hamiltonian is defined as $H=H_{I} + H_a$. Now let us write the Hamiltonian in the interaction picture with respect to $H_0$ by making the transformation  
\begin{eqnarray}
    \Tilde{H}_{int} = e^{i H_0 t} (H-H_0) e^{-i H_0 t}.
\end{eqnarray}
Further expansion of the above equation reads as follows.
\begin{widetext}
\begin{eqnarray}
    \begin{aligned}
\Tilde{H}_{int}& =\frac{\Omega_a^{(1)}}{2} e^{i(\omega_0-\omega_e+\omega_a-\delta^{(1)})t} 
\ket{0}\bra{e}+\frac{\Omega_a^{(1)}}{2} e^{-i(\omega_0-\omega_e+\omega_a -\delta^{(1)}) t} \ket{e}\bra{0} \\
& +\frac{\Omega_a^{(1)}}{2} e^{i(\omega_0-\omega_e-\omega_a-\delta^{(1)}) t}\ket{0}\bra{e}+\frac{\Omega_a^{(1)}}{2} e^{-i(\omega_0-\omega_e-\omega_a-\delta^{(1)}) t}\ket{e}\bra{0} \\
& +\frac{\Omega_b^{(1)}}{2} e^{i(\omega_1-\omega_e+\omega_b -\delta^{(1)}) t}\ket{1}\bra{e}+\frac{\Omega_b^{(1)}}{2} e^{-i(\omega_1 -\omega_e+\omega_b-\delta^{(1)}) t}\ket{e}\bra{1} \\
& +\frac{\Omega_b^{(1)}}{2} e^{i(\omega_1-\omega_e-\omega_b -\delta^{(1)})t}\ket{1}\bra{e}+\frac{\Omega_b^{(1)}}{2} e^{-i(\omega_1-\omega_e-\omega_b -\delta^{(1)}) t}\ket{e}\bra{1} \\
& +\frac{\Omega_a^{(2)}}{2} e^{i(\omega_1-\omega_{e^{\prime}}+\omega_a + \delta^{(2)}) t}\ket{1}\bra{e^\prime}+\frac{\Omega_a^{(2)}}{2} e^{-i(\omega_1-\omega_{e^\prime}+\omega_a + \delta^{(2)}) t}\ket{e^\prime}\bra{1} \\
& +\frac{\Omega_a^{(2)}}{2} e^{i(\omega_1-\omega_{e^{\prime}}-\omega_a + \delta^{(2)}) t}\ket{1}\bra{e^\prime}+\frac{\Omega_a^{(2)}}{2} e^{-i(\omega_1-\omega_{e^\prime}-\omega_a + \delta^{(2)}) t}\ket{e^\prime}\bra{1} \\
& +\frac{\Omega_b^{(2)}}{2} e^{i(\omega_0-\omega_{e^{\prime}}+\omega_b + \delta^{(2)}) t}\ket{0}\bra{e^\prime}+\frac{\Omega_b^{(2)}}{2} e^{-i(\omega_0-\omega_{e^\prime}+\omega_b + \delta^{(2)}) t}\ket{e^\prime}\bra{0} \\
& +\frac{\Omega_b^{(2)}}{2} e^{i(\omega_0-\omega_{e^{\prime}}-\omega_b + \delta^{(2)}) t}\ket{0}\bra{e^\prime}+\frac{\Omega_b^{(2)}}{2} e^{-i(\omega_0-\omega_{e^\prime}-\omega_b + \delta^{(2)}) t}\ket{e^\prime}\bra{0} \\
& -\delta^{(1)}\ket{e}\bra{e}+\delta^{(2)}\ket{e^{\prime}}\bra{e^{\prime}} .
\end{aligned}
\end{eqnarray}
\end{widetext}
Now, let us make a rotating wave approximation i.e. dropping out the fast rotating terms and simplifying the above expression by replacing $\omega_a =\omega_e -\omega_0 +\delta^{(1)} = \omega_1 -\omega_{e^\prime} +\delta^{(2)}$ and $\omega_b =\omega_e -\omega_1 +\delta^{(1)} = \omega_0 -\omega_{e^\prime} +\delta^{(2)}$. The interaction Hamiltonian reduces to 
\begin{widetext}
\begin{eqnarray}
\begin{aligned}
\Tilde{H}_{int} &=\frac{\Omega_a^{(1)}}{2} {\big[|0\rangle\bra{e}+\ket{e}\bra{0}\big]+\frac{\Omega_b^{(1)}}{2}\big[\ket{e}\bra{1}+\ket{1}\bra{e}\big] } 
+\frac{\Omega_a^{(2)}}{2}  {\big[\ket{1} \bra{e^{\prime}}+\ket{e^{\prime}}\bra{1}\big]+\frac{\Omega_b^{(2)}}{2}\big[\ket{0} \bra{e^{\prime}}+\ket{e^{\prime}}\bra{0}\big] } \\
& -\delta^{(1)}\ket{e}\bra{e}+\delta^{(2)}\ket{e^{\prime}}\bra{e^{\prime}}.
\end{aligned}
\end{eqnarray}
\end{widetext}
Let us make an adiabatic elimination of the states $\ket{e}$ and $\ket{e^\prime}$. To do so we first write $\ket{\psi} = C_0(t) \ket{0} + C_1(t) \ket{1} + C_e(t) \ket{e} + C_{e^\prime}(t) \ket{e^\prime}$. Expanding the terms of the Schr\"{o}dinger equation $-i  \frac{\partial \ket{\psi} }{\partial t} = \Tilde{H}_{int} \ket{\psi}$ and splitting the equation into relevant terms expressed below
\begin{eqnarray}
    \begin{aligned}
    \label{eq:OpticalB}
 -i \dot{C}_0(t)&=\frac{\Omega_a^{(1)}}{2} C_e(t)+\frac{\Omega_b^{(2)}}{2} C_{e^{\prime}}(t), \\
 -i \dot{C}_1(t)&=\frac{\Omega_b^{(1)}}{2} C_e(t)+\frac{\Omega_a^{(2)}}{2} C_{e^{\prime}}(t), \\
 -i \dot{C}_e(t)&=\frac{\Omega_a^{(1)}}{2} C_0(t)+\frac{\Omega_b^{(1)}}{2} C_1(t)-\delta^{(1)} C_e(t), \\
 -i \dot{C}_{e^{\prime}}(t)&=\frac{\Omega_a^{(2)}}{2} C_2(t)+\frac{\Omega_b^{(2)}}{2} C_0(t)+\delta^{(2)} C_e(t).
\end{aligned}
\end{eqnarray}
For the adiabatic elimination, we will write $\dot{C}_e (t) = 0$ and $\dot{C}_{e^ \prime} (t) = 0$. Thus we can write 
\begin{eqnarray}
    \begin{aligned}
 C_e(t) &=\frac{\Omega_a^{(1)} }{2 \delta^{(1)}} C_0(t)+\frac{\Omega_b^{(1)}}{2 \delta^{(1)}} C_1(t), \\
 C_{e^{\prime}}(t) &=-\frac{\Omega_a^{(2)}}{2 \delta^{(2)}} C_1(t)-\frac{\Omega_b^{(2)}}{2 \delta^{(2)}} C_0(t).
\end{aligned}
\end{eqnarray}
Putting these results into equation \ref{eq:OpticalB} we get 
\begin{eqnarray}
    \begin{aligned}
-i \dot{C}_0(t) & =\frac{\big[\Omega_a^{(1)}\big]^2}{4 \delta^{(1)}} C_0(t)+\frac{\Omega_a^{(1)} \Omega_b^{(1)}}{4 \delta^{(1)}} C_1(t)\\
&-\frac{\big[\Omega_b^{(2)}\big]^2}{4 \delta^{(2)}} C_0(t)-\frac{\Omega_a^{(2)} \Omega_b^{(2)}}{4 \delta^{(2)}} C_1(t), \\
-i\dot{C}_1(t) &=\frac{\big[\Omega_b^{(1)}\big]^2}{4 \delta^{(1)}} C_1(t)+\frac{\Omega_a^{(1)} \Omega_b^{(1)}}{4 \delta^{(1)}} C_0(t)\\
&-\frac{\Omega_a^{(2)} \Omega_b^{(2)}}{4 \delta^{(2)}} C_0(t)-\frac{\big[\Omega_a^{(2)}\big]^2}{4 \delta^{(2)}} C_1(t).
\end{aligned}
\end{eqnarray}
The above equation can be expressed as an effective Hamiltonian for a two-level system, i.e. the reduced form of the Schr\"{o}dinger equation reads as $\Tilde{H}_{eff} \ket{\psi_{eff}} = -i \frac{\partial \ket{\psi_{eff}}}{\partial t}$. Here $\ket{\psi_{eff}} = C_0(t) \ket{0} + C_1(t) \ket{1}$. The effective Hamiltonian can be written as 
\begin{eqnarray}
    \Tilde{H}_{eff} = \frac{\Omega_T}{2} \big[\ket{0}\bra{1} + \ket{1}\bra{0}\big] + \Delta_0 \ket{0} \bra{0} +  \Delta_1 \ket{1}\bra{1}, \nonumber \\
\end{eqnarray}
where 
\begin{eqnarray}
\begin{aligned}
    \Omega_T &= \frac{\Omega_a^{(1)} \Omega_b^{(1)}}{2 \delta^{(1)}}-\frac{\Omega_a^{(2)} \Omega_b^{(2)}}{2 \delta^{(2)}},\\
    \Delta_0 &= \frac{\big[\Omega_a^{(1)}\big]^2}{4 \delta^{(1)}}-\frac{\big[\Omega_b^{(2)}\big]^2}{4 \delta^{(2)}},\\
    \Delta_1 &= \frac{\big[\Omega_b^{(1)}\big]^2}{4 \delta^{(1)}}-\frac{\big[\Omega_a^{(2)}\big]^2}{4 \delta^{(2)}}.
    \end{aligned}
\end{eqnarray}
Here, we have the expression for the two-photon Rabi frequency shown in the main manuscript. One could also follow the same derivation for the other three transitions that we discussed in the main manuscript. In all cases, the expression for the two-photon Rabi frequency comes to be the same. The main difference will be noticeable for the AC-Stark shift and some overall signs, which are not relevant to the current discussion. In all our calculations of the two-photon Rabi frequencies, we define the detuning with respect to optical transitions, i.e. $\delta^{(1)} = \omega_a -(\omega_e-\omega_0)$ and $\delta^{(2)} = \omega_a -(\omega_1-\omega_{e^\prime})$, so while using the Rabi frequency formula one should be cautious. 

\section{Numerical simulation using time-dependent Schr\"odinger equation}
\subsection{TDSE simulation of a trapped ion experiencing micromotion and oscillating magnetic fields} \label{sec:TDSEmicro}
\begin{figure}
    \centering
\includegraphics[scale=1]{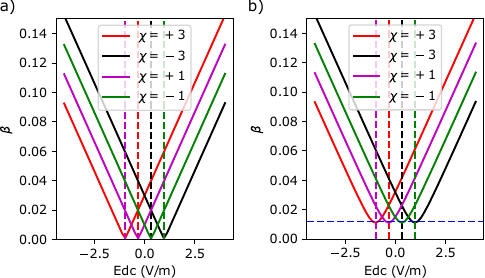}
    \caption{Simulated micromotion compensation curves for four transitions having different sensitivities to $B_{||}$, (a) when there is no phase mismatch between the RF electrodes and (b) when there is a phase mismatch which gives rise to $\beta_Q = 0.012$.}
    \label{fig:simlongmicro}
\end{figure}

A time-dependent Schr\"odinger equation (TDSE) is solved for a two-level atom using the Python-based QuTiP library \cite{johansson2012qutip}. At first, we simulate micromotion compensation curves for transitions showing differential sensitivity to $B_{||}$. The Hamiltonian is constructed from the equations described in Sec. \ref{sec:theory}. The results are presented for four transitions (Fig.~\ref{fig:simlongmicro}), similar to the experimental case described in subsection \ref{subsec:longcomp}. We see that the compensation voltages depend upon the Zeeman susceptibility $\chi$ and the value $B_{||}$. The results also confirm the offset due to phase mismatch between RF electrodes, however, this does not change the effect of $B_{||}$. 

\subsection{TDSE simulation showing effects of $B_{\perp}$ on the micromotion compensation} TDSE simulations are carried out for the transitions described in the main manuscript. The polarizations are chosen to be -70, -45, +45, and +70 degrees from the unit vector $\hat{n}$. Rabi frequencies for the involved transition are shown in Table \ref{tab:RabiDeltam2}. For the TDSE simulations, a three-level atom is considered with levels $\ket{0} = \ket{S_{1/2},m=\pm 1/2}$, $\ket{1} = \ket{D_{5/2},m=\pm5/2}$ and $\ket{e} = \ket{D_{5/2},m=\pm3/2}$ of a calcium ion. The transitions $\ket{0} \leftrightarrow \ket{1}$ and $\ket{0} \leftrightarrow \ket{e}$ take place through quadrupole transition rules with the laser fields. A magnetic dipole coupling is considered to be between $\ket{1} \leftrightarrow \ket{e}$, parameters given in Table \ref{tb:tableCoupleStrngth}. For the two-photon study, we tune the laser fields to the red/blue micromotion sideband. The coupling strength of the sideband transitions is evaluated for the cases of interest.  

\begin{table}
    \centering
    \begin{tabular}{|c|c|c|c|c|}
    \hline
       \backslashbox{Transitions}{Polarization angles} & -70\textdegree & -45\textdegree & 45\textdegree & 70\textdegree \\
    \hline
        $\ket{m=1/2}  \leftrightarrow \ket{m=5/2} $ &40 & 95 & 104 & 52 \\
    \hline
        $\ket{m=1/2}  \leftrightarrow \ket{m=3/2} $ & 119 & 95 & 86 & 114\\
    \hline
       $\ket{m=-1/2}  \leftrightarrow \ket{m=-5/2} $ & 62 & 144 & 158 &  80\\
    \hline
        $\ket{m=-1/2}  \leftrightarrow \ket{m=-3/2} $ & 181 & 121 & 130& 173\\
    \hline
    \end{tabular}
    \caption{Coupling strength $\Omega/2\pi$ (kHz) between levels separated $\Delta m = \pm 2$ (main transitions) and  levels separated by $\Delta m = \pm 1$ (mediator level). Here, the values are corrected by taking the AOM efficiency for various transitions into account. }
    \label{tab:RabiDeltam2}
\end{table}

\begin{figure}[h]
    \centering
   \includegraphics[scale=1]{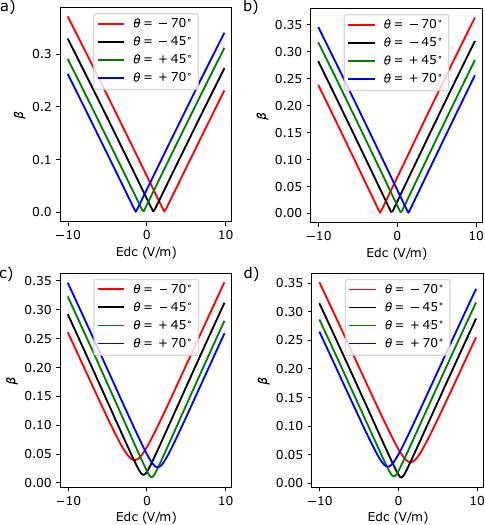} 
    \caption{Numerical simulation showing the effect of two-photon processes competing with micromotion for four polarization angles. (a)/(b) corresponds to the red/blue sideband of the $\Delta m={+2/-2}$ transition while assuming $\theta_{B_\perp} =0$. Alternatively, when we assume $\theta_{B_\perp} =2.5$~radians, the simulation results are shown in (c) and (d). Qualitatively, two sets of results show dependence on the angle $\theta_{B_\perp}$.}
    \label{fig:SimTwoPhot}
\end{figure}

The results are presented in Fig.~\ref{fig:SimTwoPhot}(a) for the red sideband of the $\Delta m=+2$ and in Fig.~\ref{fig:SimTwoPhot}(b) for the blue sideband of the $\Delta m=-2$ transitions. In the simulations, we assume that the $B_\perp$ is directed perpendicular to the laser propagation $\boldsymbol{\hat{k}}$ and the quantization field $\boldsymbol{\hat{z}}$. We here notice that the micromotion minimization is altered by the presence of the two-photon coupling. For plots (c) and (d) an angle of $\theta_{B_\perp} = 2.5$~radians is assumed. 

\subsection{Least-squares fitting of the experimental results obtained in the QSim trap}
The least-squares fitting is performed for the experimental data presented in Fig.~\ref{fig:combinedEffect} and parameters $\beta_{Q}$ and $\theta_{B_\perp}$ are estimated from the fits. For this, the TDSE is solved in QuTiP \cite{johansson2012qutip} and parameters are iteratively optimized while reducing the Chi-squared values. 

\bibliography{bibliography}
\end{document}